\documentclass[copyright,creativecommons]{eptcs}
\providecommand{\event}{HCVS 2022} 
\usepackage{breakurl}             
\usepackage{underscore}           

\usepackage{amsmath}
\usepackage{graphics}

\newtheorem{theorem}{Theorem}
\newtheorem{example}{Example}

\newcommand{\senc}{\mathsf{senc}}
\newcommand{\penc}{\mathsf{penc}}

\newcommand{\sign}{\mathsf{sign}}

\newcommand{\sdec}{\mathsf{sdec}}

\newcommand{\attacker}{\mathsf{att}}
\newcommand{\att}{\attacker}
\newcommand{\rewrite}{\Rightarrow}
\newcommand{\skA}{{\it sk}_A}

\newcommand{\pkB}{{\it pk}_B}
\newcommand{\secret}{\mathsf{s}}

\newcommand{\isnat}{\mathsf{is\_nat}}
\newcommand{\isnotnat}{\neg\mathsf{is\_nat}}

\newcommand{\kevent}{\mathbf{event}}
\newcommand{\Precise}{\mathsf{Precise}}
\newcommand{\occ}{\mathit{occ}}

\title{The Security Protocol Verifier ProVerif\\
  and its Horn Clause Resolution Algorithm}
\author{Bruno Blanchet
\institute{Inria\\ Paris, France}
\email{bruno.blanchet@inria.fr}
}
\def\titlerunning{The Security Protocol Verifier ProVerif and its Horn Clause Resolution Algorithm}
\def\authorrunning{B. Blanchet}
\begin{document}
\maketitle

\begin{abstract}
ProVerif is a widely used security protocol verifier. Internally,
ProVerif uses an abstract representation of the protocol by Horn
clauses and a resolution algorithm on these clauses, in order to prove
security properties of the protocol or to find attacks.  In this
paper, we present an overview of ProVerif and discuss some
specificities of its resolution algorithm, related to the particular
application domain and the particular clauses that ProVerif
generates. This paper is a short summary that gives pointers to
publications on ProVerif in which the reader will find more details.
\end{abstract}

\section{Introduction}

The verification of security protocols is a very active research area since the 1990's. Security protocols are ubiquitous: Internet (in particular, the TLS protocol used for \texttt{https://} connections), WiFi, mobile phones, credit cards, \dots. Their design is notoriously error-prone, and errors are not detected by testing: they appear only when an adversary tries to attack the protocol. Therefore, it is important to formally verify them.

In order to formalize security protocols, one needs a mathematical model for them. One typically considers an active adversary, which can listen to messages sent on the network, compute its own messages, and send them to the network as if they came from honest participants. To facilitate the automatic verification of protocols, most protocol verifiers consider the symbolic model of cryptography, also called ``Dolev-Yao model''~\cite{Needham78,Dolev83}. In this model, cryptographic primitives, such as encryption, are considered as ideal black-boxes, represented by function symbols; messages are modeled by terms on these primitives; and the adversary is restricted to apply defined primitives. This is also called the perfect cryptography assumption: the only way for the adversary to decrypt a message is to use the decryption function with the correct key. In such a model, one of the main tasks of protocol verification consists in computing the knowledge of the adversary, that is, the set of terms that the adversary can obtain. This is still non-trivial as this set is typically infinite, but it is much simpler than reasoning about bitstrings and probabilities as in cryptographic proofs. The two most widely used symbolic protocol verifiers are probably ProVerif~\cite{Blanchet16b} and Tamarin~\cite{Meier13}. For more details on the field of protocol verification, we refer the reader to the surveys~\cite{BlanchetETAPS12,BarbosaetalOakland21}.
In this paper, we focus on the protocol verifier ProVerif, which can be downloaded from \url{https://proverif.inria.fr}. We present an overview of ProVerif in the next section and focus on its Horn clause resolution algorithm in Section~\ref{sec:resol}.

\section{ProVerif}

\begin{figure}[t]
  \begin{center}
    \input{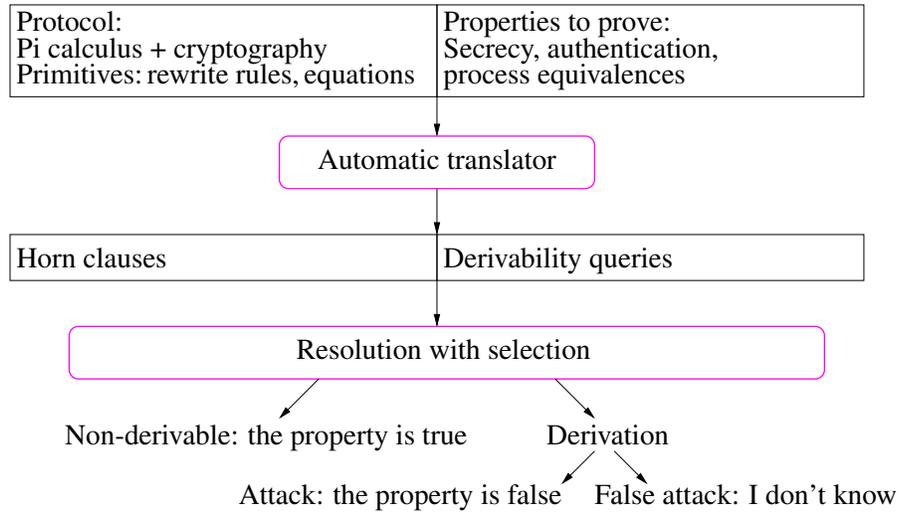}
  \end{center}
  \caption{The structure of ProVerif}\label{fig:PVstructure}
\end{figure}

As illustrated in Figure~\ref{fig:PVstructure}, ProVerif takes as input a description of the protocol to verify in an extension of the pi calculus with function symbols to represent cryptography, a dialect of the applied pi calculus~\cite{Abadi2001,AbadiBlanchetFournetJACM17}. This language can be seen as a small, domain-specific, programming language for security protocols. Additionally, one needs to define the cryptographic primitives, using rewrite rules or equations. For instance, shared-key encryption can be represented by a free function symbol $\senc$ that takes as argument a message and a key and returns the corresponding ciphertext. The corresponding decryption function $\sdec$ can be defined by a rewrite rule
\[\sdec(\senc(m, k), k) \rightarrow m\]
which means that, when we decrypt a ciphertext $\senc(m, k)$ with the correct key $k$, we obtain the cleartext $m$.
Free function symbols like $\senc$ are named \emph{constructors} while function symbols defined by rewrite rules are named \emph{destructors}.

ProVerif also takes as input the security properties to prove, named queries, which can be:
\begin{itemize}
\item \emph{secrecy} properties~\cite{Abadi04c}: the adversary cannot compute certain values;
\item \emph{authentication} properties~\cite{Blanchet08c}: if some participant Alice thinks she talks to Bob, then she really talks to Bob. Authentication is formalized by correspondence properties~\cite{Woo93}, of the form: if some event has been executed, then some other event has been executed. Events can represent that Alice concluded the protocol apparently with Bob, or that Bob started the protocol, apparently with Alice.
\item \emph{equivalence} properties~\cite{Blanchet07b}: the adversary cannot distinguish two protocols. Equivalence properties are a powerful notion to specify security properties (such as anonymity and privacy), but they are also difficult to verify. ProVerif can verify only a strong notion of equivalence, named \emph{diff-equivalence}, between protocols that have the same structure but differ only by the messages they exchange.
\end{itemize}

ProVerif proves such security properties thanks to an internal representation of the protocol by Horn clauses, obtained by an automatic translation from the protocol and the cryptographic primitives. Simplifying as much as possible, the clauses use as main predicate $\att$: $\att(M)$ means that the adversary may know the term $M$. Some clauses represent computations by the adversary:
\begin{itemize}
\item For each constructor $f$ of arity $n$, the clause
  \[\attacker(x_1) \wedge \ldots \wedge \attacker(x_n) \rewrite
  \attacker(f(x_1, \ldots, x_n))\]
  is generated, representing that the adversary can compute $f(x_1, \ldots, x_n)$ by applying $f$ when it has $x_1$, \dots, $x_n$.
  For instance, for shared-key encryption $\senc$, the following clause is generated:
  \begin{equation}
    \attacker(m) \wedge \attacker(k) \rewrite \attacker(\senc(m,k))\tag{$\senc$}\label{cl:senc}
  \end{equation}
\item For each destructor $g$, defined by a rewrite rule  $g(M_1, \ldots, M_n) \rightarrow M$, the clause 
  \[\attacker(M_1) \wedge \ldots \wedge \attacker(M_n) \rewrite
  \attacker(M)\]
  is generated, representing that the adversary can compute $M$ when it has $M_1, \dots, M_n$, by applying $g$.
  For instance, for shared-key decryption $\sdec$, defined by $\sdec(\senc(m, k), k) \rightarrow m$, the following clause is generated:
  \begin{equation}
    \attacker(\senc(m, k)) \wedge \attacker(k) \rewrite \attacker(m)\tag{$\sdec$}\label{cl:sdec}
  \end{equation}
  If the adversary has the ciphertext $\senc(m, k)$ and key $k$, it can obtain the cleartext $m$ by decryption.
  
\end{itemize}
Other clauses represent the protocol itself:
if a principal $A$ has received the messages $M_1, \ldots, M_n$
and sends the message $M$, the following clause is generated:
\[\attacker(M_1) \wedge \ldots \wedge \attacker(M_n) \rewrite \attacker(M)\,.\]
Indeed, if the adversary has $M_1$, \dots, $M_n$, it can send them to $A$,
which is going to reply with $M$. The adversary intercepts this message and thus obtains $M$.
For instance, consider the following protocol, inspired by the
Denning-Sacco key distribution protocol~\cite{Denning81}:
\begin{center}
\begin{tabular}{l l}
Message 1. & $A \rightarrow B : \penc(\sign(k,\skA),\pkB)$\\
Message 2. & $B \rightarrow A : \senc(\secret, k)$\\
\end{tabular}
\end{center}
The symbol $\penc$ represents public-key encryption,
$\sign$ represents a signature. 
In this protocol, $A$ generates a fresh key $k$ and aims to share
it with $B$. $A$ signs $k$ with her secret key $\skA$ and
encrypts it under $B$'s public key $\pkB$. Only $B$ can decrypt this message;
then $B$ verifies $A$'s signature and obtains the key $k$.
In the second message, $B$ uses this key $k$ to encrypt a secret $\secret$
under $k$, using shared-key encryption.

In this protocol, upon receipt of a message of the form $\penc(\sign(y,\skA),\pkB)$, $B$ replies with the message $\senc(\secret, y)$, so the generated clauses include
\[\attacker(\penc(\sign(y,\skA),\pkB)) \rewrite \attacker(\senc(\secret, y))\,.\]
This clause represents that the adversary sends
$\penc(\sign(y,\skA),\pkB)$ to $B$, and intercepts his reply
$\senc(\secret, y)$.

ProVerif also translates the security properties to prove into derivability queries on the generated Horn clauses. For instance, in order to prove that $\secret$ is secret, ProVerif proves that $\att(\secret)$ is \emph{not} derivable from the generated Horn clauses. Intuitively, the adversary is then unable to compute $\secret$. More generally, a security property is proved when no instance of a certain fact $F$ is derivable from the clauses. ProVerif uses a resolution algorithm (detailed in the next section) in order to determine whether some instance of $F$ is derivable from the clauses or not. 

When an instance of $F$ is derivable from the clauses, the derivation is the witness of an attack. However, the Horn clause representation introduces an abstraction: mainly, the Horn clauses can be applied any number of times in a derivation~\cite{Blanchet05b}, but that does not always correspond to what happens in the protocol, since some protocol steps may be applicable only once, for instance. Because of this abstraction, a derivation, that is, an attack at the Horn clause level, does not always correspond to an attack at the protocol level. ProVerif therefore uses an attack reconstruction algorithm~\cite{Allamigeon05} in order to reconstruct an attack at the protocol level from the derivation. When this algorithm succeeds in finding an attack, the security property is definitely false. When this attack reconstruction algorithm fails, the derivation is a so-called ``\emph{false attack}''; in this case, we do not know whether the security property holds or not (see Example~\ref{ex:precise1} below for an example).

As Horn clauses can be used for resolution an unbounded number of times by default, this abstraction allows ProVerif to prove properties for an unbounded number of sessions of the protocol, an undecidable problem~\cite{Durgin04}. However, because of this abstraction, ProVerif is not complete: it does not always succeed in deciding whether a security property holds or not. Moreover, in general, the resolution algorithm may not terminate. In practice, ProVerif is still precise and efficient for many examples of protocols.

\section{The resolution algorithm}\label{sec:resol}

The resolution algorithm of ProVerif is based on resolution with free selection~\cite{Bachmair01}. A selection function selects one literal in each clause, and the algorithm performs resolution upon selected literals, that is, from two clauses $R = H \rewrite C$ and $R' = F' \wedge H' \rewrite C'$ where the conclusion $C$ is selected in $R$ and the hypothesis $F'$ is selected in $R'$, the algorithm generates the clause $H\sigma \wedge H'\sigma \rewrite C'\sigma$, where $\sigma$ is the most general unifier of $C$ and $F'$. Intuitively, this clause is obtained by using $R$ to infer $C \sigma = F'\sigma$ from $H\sigma$, and then using $R'$ to infer $C'\sigma$ from $F'\sigma$ and $H'\sigma$. These resolution steps are performed between all clauses until a fixpoint is reached, that is, no new clause can be added. Finally, among the clauses in this fixpoint, only the clauses in which the conclusion is selected are kept.

The following theorem states the soundness and completeness of resolution with free selection.
It is a particular case of~\cite[Lemma~2]{Blanchet08c}.

\begin{theorem}
The set of clauses obtained by resolution with free selection derives the same facts as the initial clauses.
\end{theorem}

The key idea is to choose the selection function to avoid resolving upon facts of the form $\att(x)$ for a variable $x$, because such facts unify with any fact $\att(M)$, yielding many resolution steps and non-termination. Hence a basic selection function selects some hypothesis not of the form $\att(x)$ if possible, and the conclusion when all hypotheses are of the form $\att(x)$.

ProVerif uses standard optimizations of resolution provers (elimination of subsumed clauses, of tautologies, \dots)~\cite{Blanchet08c}. We do not detail those further. However, it also uses less standard, domain-specific optimizations and extensions. We sketch a few of them below, with references of papers in which more details can be found.

\paragraph{Data constructors~\cite{Blanchet08c}}
Data constructors are constructors $f$ that come with associated projections $\pi^f_i$ defined by $\pi^f_i(f(x_1, \dots,x_n)) = x_i$ for $1 \leq i \leq n$. Such constructors model data structures that appear in protocols; projections allow to obtain the elements of the structure. For such constructors, we have clauses
\begin{align}
  &\attacker(x_1) \wedge \ldots \wedge \attacker(x_n) \rewrite
  \attacker(f(x_1, \ldots, x_n))  \tag{for $f$}\\
  &\attacker(f(x_1, \ldots, x_n)) \rewrite \attacker(x_i) \tag{for $\pi^f_i$}
\end{align}
Therefore, $\attacker(f(x_1, \ldots, x_n))$ is equivalent to $\attacker(x_1) \wedge \ldots \wedge \attacker(x_n)$. So we can simplify clauses by decomposing data constructors, except in the two clauses above: we replace $\attacker(f(x_1, \ldots, x_n))$ with $\attacker(x_1) \wedge \ldots \wedge \attacker(x_n)$ in hypotheses of clauses and we replace clauses $H \rewrite \attacker(f(x_1, \ldots, x_n))$ with $n$ clauses $H \rewrite \attacker(x_i)$.
(This transformation corresponds to resolving each clause $R$ with the clauses for $f$ and for $\pi^f_i$. However, the difference with an ordinary resolution step is that the initial clause $R$ can be removed. Only the transformed clause is kept.)

\paragraph{Blocking predicates~\cite{Blanchet08c}}
Blocking predicates are predicates that occur in the hypothesis of clauses and on which we do not resolve: the selection function never selects them. Hence, they remain in the hypothesis of clauses until the end of the resolution algorithm. ProVerif proofs are valid for any definition of these predicates. For instance, they can be used to model conditions that could not be explicitly defined in ProVerif, such as conditions on real numbers. They are also very useful in order to prove correspondence properties, as explained in~\cite[Section~4]{Blanchet08c}: we represent events that we wish to prove using a blocking predicate; the presence of these blocking events in the hypotheses of clauses at the end of resolution shows that they must have been executed in order to reach the conclusion of these clauses.

\paragraph{Disequations~\cite{Blanchet07b}}
ProVerif also supports disequality constraints modulo the equations that define the cryptographic primitives, of the form $\forall x_1, \dots, x_n. M \neq N$. These disequality constraints may occur in the hypothesis of clauses. They are handled in the resolution algorithm using simplifications explained in~\cite{Blanchet07b}.

\paragraph{Natural numbers~\cite{Blanchet22}}
ProVerif supports natural numbers, which can be used to represent counters: it supports constant natural numbers, addition and subtraction of a variable and a constant, comparisons, and tests whether a term is a natural number or not. Natural numbers are implemented in clauses with constraints  $\isnat(M)$ ($M$ is a natural number), $\isnotnat(M)$ ($M$ is not a natural number), and $M \geq N + n$ where $n$ is a constant natural number. The latter constraints are simplified using the Bellman-Ford algorithm~\cite{Bellman58}.

\paragraph{Temporal correspondence queries~\cite{Blanchet22}}
ProVerif supports correspondence queries that use events of the form $\kevent(M)@i$, where $i$ is a temporal variable: $\kevent(M)@i$ means that event $M$ happened at step $i$. One can then compare temporal variables with each other: for instance $\kevent(A(x))@i \mathrel{\&\&} \kevent(B(x))@j \Longrightarrow i < j$ means that if event $A(x)$ happens at step $i$ and $B(x)$ happens at step $j$ then $i < j$, so event $A(x)$ happens before event $B(x)$. These queries are encoded using special natural number constraints $i < j$ and $i \leq j$.

\paragraph{Precise actions~\cite{Blanchet22}}
Precise actions allow us to avoid false attacks that come from the repeated usage of Horn clauses in a derivation, as illustrated by the following example.

\begin{example}\label{ex:precise1}
  Consider a protocol in which the participant $A$ sends the messages $\senc(k_1,k)$, $\senc(k_2,k)$, and $\senc(s, (k_1,k_2))$, 
  and the participant $B$ receives one message $x$ and sends its decryption under $k$, where $k$, $k_1$, $k_2$ are keys initially secret and $s$ is a secret. $B$ acts as a decryption oracle under $k$, but only once. Therefore, the adversary can obtain either $k_1$ or $k_2$ by decrypting either $\senc(k_1,k)$ or $\senc(k_2,k)$ using $B$, but it cannot obtain both $k_1$ and $k_2$, hence it cannot decrypt $\senc(s, (k_1,k_2))$ and cannot obtain~$s$.

  The clauses for this protocol include:
  \begin{align}
    &\attacker(\senc(k_1,k))\qquad
    \attacker(\senc(k_2,k))\qquad
    \attacker(\senc(s,(k_1,k_2)))\tag*{for $A$}\\
    &\attacker(\senc(y,k)) \rewrite \attacker(y)\tag*{for $B$}\\
    &\attacker(x) \wedge \attacker(y) \rewrite \attacker((x,y))\tag*{for the pair}
  \end{align}
  as well as the clauses~\eqref{cl:senc} and~\eqref{cl:sdec} for
  encryption and decryption.
  The clause for $B$ can be applied any number of times in a derivation. Using this clause, we derive $\attacker(k_1)$ from $\attacker(\senc(k_1,k))$ and
  $\attacker(k_2)$ from $\attacker(\senc(k_2,k))$. Then we derive $\attacker((k_1,k_2))$ by the clause for the pair, and $\attacker(s)$ using $\attacker(\senc(s,(k_1,k_2)))$ and the clause for $\sdec$. Therefore, at the Horn clause level, the adversary can obtain $s$: this is a false attack, due to the repeated usage of the clause for $B$, which does not match the specification of the protocol. With these clauses, ProVerif is unable to prove secrecy of $s$.
  This false attack can be avoided by tagging \texttt{precise} the input that receives message $x$ in $B$.
\end{example}
Precise actions are implemented as a particular axiom, as we explain below.

\paragraph{Restrictions, axioms, lemmas~\cite{Blanchet22}}
Syntactically, restrictions, axioms, and lemmas are particular correspondence queries. However, they play a different role:
\begin{itemize}
\item Restrictions restrict the set of traces on which queries are proved: queries are proved only on traces that satisfy the restrictions.
\item Axioms are properties that hold on all considered traces, but are not proved by ProVerif: they are assumed. For instance, they are useful to use properties that are proved manually and that ProVerif cannot prove.
\item Lemmas are proved by ProVerif, and then assumed in the proof of subsequent lemmas and queries.
\end{itemize}
ProVerif uses restrictions, axioms, and already proved lemmas in order to strengthen clauses generated during resolution, as follows. Suppose that a proved lemma (or axiom or restriction) is $\bigwedge_{i} F_i \Longrightarrow \phi$ and that resolution generates a clause $H \rewrite C$ such that for all $i$, $F_i\sigma \in H$ or $F_i\sigma = C$. Then we know that $\bigwedge_{i} F_i\sigma$ holds, hence by the lemma, $\phi\sigma$ holds as well. Therefore, we add $\phi\sigma$ in the hypothesis of the clause, yielding $H \wedge \phi\sigma \rewrite C$.

\begin{example}[Example~\ref{ex:precise1} continued]
  Precise actions are implemented using axioms. After each precise input that receives message $x$, we generate a fresh name $\occ$ and execute the event $\Precise(\occ,x)$, and we add the axiom
  \begin{equation}
    \kevent(\Precise(\occ,x_1)) \mathrel{\&\&} \kevent(\Precise(\occ,x_2)) \Longrightarrow x_1 = x_2
    \tag{$\Precise$}\label{ax:precise}
  \end{equation}
  This axiom means that, if events $\Precise(\occ,x_1)$ and $\Precise(\occ,x_2)$ are executed with the same value of $\occ$ (hence after the same execution of the precise input), then $x_1 = x_2$, that is, the received message $x$ is the same.

  In Example~\ref{ex:precise1}, there is a single execution of the precise input, the input of $B$, so a single name $\occ$. Therefore, the axiom shows that a single message $x$ can be processed by $B$.

  During resolution with the precise input, the following clause is generated
  \[\kevent(\Precise(\occ,\senc(k_1,k))) \wedge \kevent(\Precise(\occ,\senc(k_2,k))) \rewrite \attacker(s)\]
  which means that if $B$ has been used to decrypt $\senc(k_1,k)$ (event $\Precise(\occ,\senc(k_1,k))$ has been executed) and $\senc(k_2,k)$ (event $\Precise(\occ,\senc(k_2,k))$ has been executed), then the adversary obtains $s$.
  Using the axiom~\eqref{ax:precise}, this clause is transformed into
  \[\kevent(\Precise(\occ,\senc(k_1,k))) \wedge \kevent(\Precise(\occ,\senc(k_2,k))) \wedge \senc(k_1,k) = \senc(k_2,k) \rewrite \attacker(s)\]
  and hence removed because $\senc(k_1,k) \neq \senc(k_2,k)$.
  That avoids the false attack that we had previously and ProVerif can now prove the secrecy of $s$.
\end{example}

\paragraph{Proofs by induction~\cite{Blanchet22}}
In order to perform proofs by induction, we use as lemma the property that we want to prove itself, but on a strict prefix of the trace. Since intuitively the hypothesis of a clause happens strictly before the conclusion, when we apply an inductive lemma to a clause, its hypothesis must be proved using only the hypothesis of the clause, not the conclusion of the clause.

\bigskip
All extensions of~\cite{Blanchet22} are recent extensions of ProVerif designed and implemented mainly by Vincent Cheval. As explained in~\cite[Section~4]{Blanchet22}, he also optimized ProVerif considerably, by revising many of its algorithms. In particular, using Schulz's idea of feature vertex indexing~\cite{Schulz13} allowed him to speed up subsumption tests considerably and indexing clauses in a prefix tree allowed him to speed up the resolution steps themselves.

\section{Conclusion}

Horn clauses are a powerful mechanism to reason about terms, hence
about security protocols in the symbolic model. As shown in
Section~\ref{sec:resol}, tuning the resolution algorithm with specific
clause simplifications and transformations allows one to implement
many optimizations and extensions that are essential for reasoning
about protocols.

\paragraph{Acknowledgment}
This work was partly supported by ANR TECAP (decision number ANR-17-CE39-0004-03).


\end{document}